\documentclass[a4paper, 11pt]{amsart}
\usepackage{graphicx}
\usepackage{bm}
\usepackage{amsmath}
\usepackage{amsfonts}
\usepackage{amssymb}

\begin{document}

\title{Character varieties and algebraic surfaces for the topology of quantum computing}

\author{Michel Planat$\dag$, David Chester$\ddag$, Raymond Aschheim$\ddag$, Marcelo M. Amaral$\ddag$, Fang Fang$\ddag$ and Klee Irwin$\ddag$}

\address{$\dag$ Universit\'e de Bourgogne/Franche-Comt\'e, Institut FEMTO-ST CNRS UMR 6174, 15 B Avenue des Montboucons, F-25044 Besan\c con, France.}
\email{michel.planat@femto-st.fr}

\address{$\ddag$ Quantum Gravity Research, Los Angeles, CA 90290, USA}
\email{davidc@QuantumGravityResearch.org}
\email{raymond@QuantumGravityResearch.org}
\email{Klee@quantumgravityresearch.org}
\email{Marcelo@quantumgravityresearch.org}
\email{Fang@QuantumGravityResearch.org}

\begin{abstract}
 
It is shown that the representation theory of some finitely presented groups thanks to their $SL_2(\mathbb{C})$ character variety is related to algebraic surfaces. 
We make use of the Enriques-Kodaira classification of algebraic surfaces and the related topological tools to make such surfaces explicit. We study the connection of $SL_2(\mathbb{C})$ character varieties to topological quantum computing (TQC) as an alternative to the concept of anyons. The Hopf link $H$, whose character variety is a Del Pezzo surface $f_H$ (the trace of the commutator), is the kernel of our view of TQC. Qutrit and two-qubit magic state computing, derived from the trefoil knot in our previous work, may be seen as TQC from the Hopf link. The character variety of some two-generator Bianchi groups as well as that of the fundamental group for the singular fibers $\tilde{E}_6$ and $\tilde{D}_4$  contain $f_H$. A surface birationally equivalent to a $K_3$ surface is another compound of their character varieties.

\end{abstract}

\maketitle






\normalsize

\section{Introduction}
 
Let $M$ be a $3$- or a $4$-manifold. An important invariant of $M$ is the fundamental group $\pi_1(M)$. It classifies the equivalence classes under homotopy of the loops contained in $M$. If $M$ is the complement of a knot (or a link) embedded in the $3$-dimensional space, $\pi_1(M)$ is called a knot group. The Wirtinger representation explicitly describes the knot group with generators and relations based on a diagram of the knot.

In Reference \cite{CullerShalen1983}, the authors introduce a technique for describing all representations of a finitely-presented group $\Gamma$ in the group $SL_2(\mathbb{C})$.
Representations of $\Gamma$ in $SL_2(\mathbb{C})$ are homomorphisms $\rho: \Gamma \rightarrow SL_2(\mathbb{C})$. The character of a representation $\rho$ is a map $\kappa_{\rho}: \Gamma \rightarrow \mathbb{C}$ defined by $\kappa_{\rho}(\gamma)=\mbox{tr}(\rho(\gamma))$, $\gamma \in \Gamma$. The set of characters of representations $\Gamma$ in $SL_2(\mathbb{C})$ is $R(\Gamma)=\mbox{Hom}(\Gamma,SL_2(\mathbb{C}) )$ which is a complex affine algebraic set.  The set of characters is defined to be $X(\Gamma)=\left\{\kappa_{\rho}|\rho \in R(\Gamma)\right\}$.

Given a manifold $M$ with fundamental group $\Gamma=\pi_1(M)$, one refers to the affine algebraic set $\tilde{X}(\pi_1(M))$ as the character variety of $M$. The character varieties of some $3$-manifolds of the Bianchi type have been investigated in \cite{Harada2012} and more generally in \cite{Ashley2017}. In the latter reference, a Sage software is developed to make the character variety explicit \cite{CharVar}. 

In the present paper, one finds that such character varieties decompose into algebraic surfaces that can be recognized through the Enriques-Kodaira classification \cite{Kodaira}. Such surfaces are candidates  for a new type of topological quantum computing different from anyons \cite{TQCanyon}. Related ideas are in References \cite{Asselmeyer1,Asselmeyer2}. A previous work of our group \cite{MPQGR1,PlanatMagic} proposed to relate the fundamental group of some $3$-manifolds to quantum computing but did not employ the representation theory.

In Section \ref{prol}, we introduce our mathematical concepts about algebraic surfaces, $SL_2(\mathbb{C})$ character varieties and magic state quantum computing. The Hopf link $H$ and the $3$-dimensional surface $f_H(x,y,z)=xyz-x^2-y^2-z^2+4$ ly at the \lq basement'. In Section \ref{TQComputing}, we investigate the \lq floors' starting with the Whitehead link and its cousins in the Bianchi group family. 
Then, we find that the affine $E_6$ manifold -- the two-covering of the affine $E_8$ manifold (also the $0$-surgery on the trefoil knot)-- and affine $D_4$ manifold --a three-covering-- are other floors upon H and $f_H$. The $SL_2(\mathbb{C})$ character variety of $\tilde{E}_6$ is made of two $K_3$ surfaces in addition to $f_H$. In conclusion, we propose a few vistas for future research.

\section{Prolegomena}
\label{prol}
\subsection{Algebraic surfaces}

Given an ordinary projective surface $S$ in the projective space $P^3$ over a number field, if $S$ is birationally equivalent to a rational surface, the software Magma \cite{Magma} determines the map to such a rational surface  and returns its type within five categories. The returned type of $S$ is $P^2$ for the projective plane, a quadric surface (for a degree $2$ surface in $P^3$), a rational ruled surface, a conic bundle or a degree $p$ Del Pezzo surface where $1 \le p \le 9$.

A further classification may be obtained for $S$ in $P^3$ if $S$ has at most point singularities. Magma computes the type of $S$ (or rather, the type of the non-singular projective surfaces in its birational equivalence class) according to the classification of Kodaira and Enriques \cite{Kodaira}. The first returned value is the Kodaira dimension of $S$, which is $-\infty$, $0$, $1$ or $2$. The second returned value further specifies the type within the Kodaira dimension $-\infty$ or $0$ cases (and is irrelevant in the other two cases).

Kodaira dimension $-\infty$ corresponds to birationally ruled surfaces. The second return in this case is the irregularity $q \ge 0$ of $S$. So $S$ is birationally equivalent to a ruled surface over a smooth curve of genus $q$ and is a rational surface if and only if $q$ is zero.

Kodaira dimension $0$ corresponds to surfaces which are birationally equivalent to a $K_3$ surface, an Enriques surface, a torus or a bi-elliptic surface.

Every surface of Kodaira dimension $1$ is an elliptic surface (or a quasi-elliptic surface in characteristics $2$ or $3$), but the converse is not true: an elliptic surface can have Kodaira dimension $-\infty$, $0$ or $1$.

Surfaces of Kodaira dimension $2$  are algebraic surfaces of general type.

\subsection{The Hopf link}
\label{HopfL}

\begin{figure}[h]
\includegraphics[width=6cm]{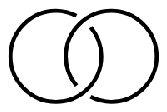}
\includegraphics[width=6cm]{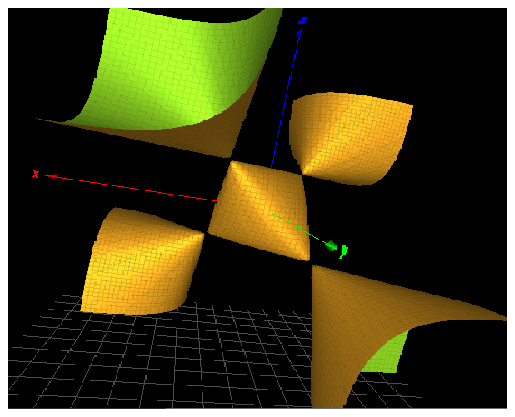}
\caption{Left: the Hopf link. Right: a $3$-dimensional picture of the $SL_2(\mathbb{C})$ character variety $\Sigma_H$ for the Hopf link complement. }
\label{Hopf}
 \end{figure}

Let us anticipate the details of our approach of connecting knot/link theory, algebraic surfaces and topological quantum computing. One takes the linking of two unknotted curves as in Fig. \ref{Hopf} (Left), the obtained link is called the Hopf link $H$=L2a1 whose knot group is defined as the fundamental group of the knot complement in the $3$-sphere $S^3$
\begin{equation}
\pi_1(S^3\setminus L2a1)=\left\langle a,b|[a,b]\right\rangle=\mathbb{Z}^2,
\label{Fricke}
\end{equation}
\noindent
where $[a,b]=abAB$ (with $A=a^{-1},~ B=b^{-1}$) is the group theoretical commutator.

They are interesting properties of the knot group $\pi_1$ of Hopf link that we would like to mention.

First, the number of coverings of degree $d$ of $\pi_1$ (which is also the number of conjugacy classes of index $d$) is precisely the sum of divisor function $\sigma(d)$ \cite{Liskovets2009}.

Second, there exists an invariance of $\pi_1$ under a repetitive action of the Golden mean substitution (the Fibonacci map) $\rho:a \rightarrow ab$, $b \rightarrow a$ or under the Silver mean substitution $\rho:a \rightarrow aba$, $b \rightarrow a$. The terms Golden and Silver refer to the Perron-Frobenius eigenvalue of the substitution matrix \cite[Examples 4.5 and 4.6]{BaakeBook}.
Such an observation links the Hopf link, the group $\pi_1$ of the $2$-torus and aperiodic substitutions.

Using Sage software \cite{CharVar} developed from Ref. \cite{Ashley2017}, the $SL_2(\mathbb{C})$ character variety is the polynomial 

\begin{equation}
f_H(x,y,z)=xyz-x^2-y^2-z^2+4.
\end{equation}

As expected, the $3$-dimensional surface $\Sigma:f_H(x,y,z)=0$ is the trace of the commutator and is known to correspond to the reducible representations \cite[Theorem 3.4.1]{Goldman2009}. A picture is given in Figure \ref{Hopf} (Right).

If we adopt the perspective of algebraic geometry by viewing $\Sigma$ in the $3$-dimensional projective space as 

\begin{equation}
\Sigma_H(x,y,z,t): xyz-t(x^2+y^2+z^2)+4 t^3=0, 
\end{equation}

\noindent
then $\Sigma_H$ is a rational surface, more precisely a degree $3$ Del Pezzo surface. It contains $4$ simple singularities.

In Reference \cite{Asselmeyer1}, the author proposes  the representation $\pi_1 \rightarrow SU(2) \otimes SU(2)$ as a model of $2$-qubit quantum computing in which each factor is associated to a single qubit located on each component of the Hopf link. Our project expands this idea by taking the representation $\pi_1 \rightarrow SL(2,\mathbb{C})$ and the attached character variety $\Sigma_H$ as a model of topological quantum computing. Ideas in this direction are found in \cite{Torsten2019}.

\subsection{Magic state quantum computing}
\label{magicQC}

Since 2017, following the seminal paper \cite{Bravyi2005}, we develop a type of universal quantum computing based on magic states \cite{PlanatGedik, Planat2018}. A magic state is a non-stabilizer pure state (a non-eigenstate of a Pauli group gate) that adds to stabilizer operations (Clifford group unitaries, preparations, and measurements) in order to ensure the universality (the possibility of getting an arbitrary quantum gate). It has been recognized that some of the magic states are fiducial states for building a minimal informationally complete positive operator valued  measure (or MIC) of the corresponding Hilbert space dimension $d$, based on the action of the Pauli group $\mathcal{P}_d$ on the state.

The lower dimensional case is the qutrit MIC arising from the fiducial state $f_{QT}=(0,1,\pm 1)$. The next case is the two-qubit MIC arising from the fiducial state $f_{2QB}=(0,1,-\omega_6,\omega_6-1)$ with $\omega_6=\exp (\frac{2i \pi}{6})$. For such magic/fiducial states, the geometry of triple products of projectors $\Pi_i=\left|\psi_i\right\rangle\left\langle  \psi_i\right|$ built with the $d^2$ outcomes $\psi_i$ is the Hesse configuration (for qutrits: in dimension $d=3$) and the $GQ(2,2)$ configuration (for two-qubits: in dimension $d=2^2$) \cite{PlanatGedik}. The latter configuration embeds the celebrated Mermin square configuration (a $3 \times 3$ grid of observables) needed to prove the Kochen-Specker theorem.

Our search of the magic states is performed with a two-generator infinite group $G$. A coset table over a subgroup $H$ of index $d$ is built by means of the Coxeter-Todd algorithm resulting in a permutation group. The latter may be seen as a $d\times d$ permutation matrix whose eigenstates are the candidates for a magic (and fiducial) state. We are dealing with low $d$ values so that the choice is not large and many groups $G$ do the job. Let us take $G$ as the modular group $\Gamma=PSL(2,\mathbb{Z})$ as in \cite{Planat2018}. Then the appropriate subgroups are in the family of congruence subgroups $\Gamma_0(N)$ of level $N$ defined as the subgroups of upper triangular matrices with entries taken modulo $N$. The index of $\Gamma_0(N)$ is the Dedekind function $\psi(N)$. Thus our relevant group for the qutrit magic state is the congruence subgroup $\Gamma_0(2)$  and for the two-qubit magic state it is $\Gamma_0(3)$.

Next, the two groups of interest $\Gamma_0(2)$  and $\Gamma_0(3)$ may be seen as fundamental groups of non-hyperbolic $3$-manifolds \cite{MPQGR1}. It is known that $\Gamma$ is isomorphic to the fundamental group $\pi_1(S^3 \setminus3_1)$ of the trefoil knot complement. The subgroups of interest of $\pi_1(S^3 \setminus3_1)\cong \Gamma$ are attached to links $L7n1$ and $L6a3$ \cite[Table 4]{PlanatMagic}. Their presentation is as follows

\begin{eqnarray}
& \pi_1(S^3 \setminus L7n1)\cong \Gamma_0(2)=\left\langle a,b | [a,b^2]\right\rangle,\nonumber\\
& \pi_1(S^3 \setminus L6a3)\cong \Gamma_0(3)=\left\langle a,b|[a,b^3]\right\rangle.\nonumber
\end{eqnarray}

Last but not least, in view of the presentation of the groups as commutators, the $SL_2(\mathbb{C})$ character variety of the two groups is that $f_H(x,y,z)$ of the Hopf link (apart for trivial factors $y$ and $y^2-1$) as shown in Table \ref{Canonical}.
One concludes that universal quantum computing based on the magic states $f_{QT}$ and $f_{2QB}$ is essentially topological quantum computing over the Hopf link. The underlying algebraic geometry for these models is the surface drawn in Figure \ref{Hopf}.

\section{Character varieties for fundamental groups of $3$-manifolds and the related algebraic surfaces} 
\label{TQComputing}

\subsection{The $SL_2(\mathbb{C})$ character varieties of knot groups whose reducible component is that of the Hopf link}

\begin{figure}[h]
\includegraphics[width=6cm]{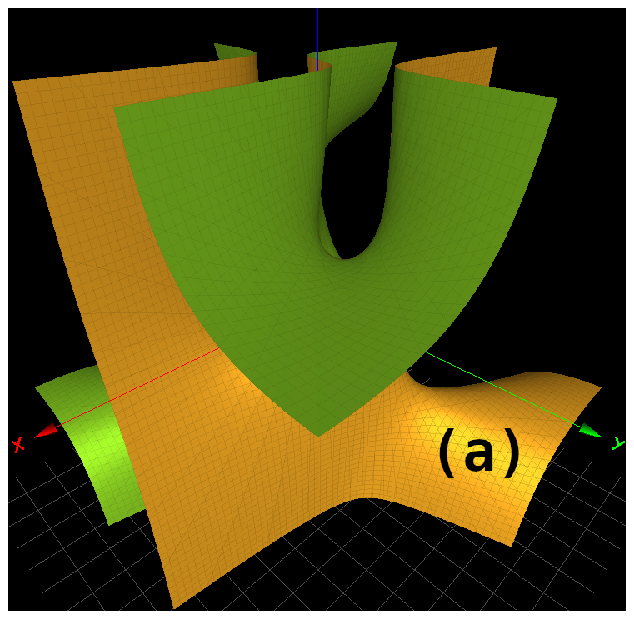}
\includegraphics[width=6cm]{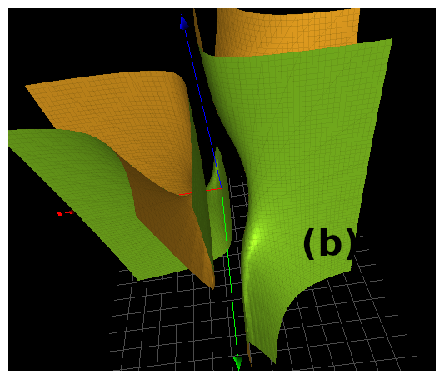}
\includegraphics[width=6cm]{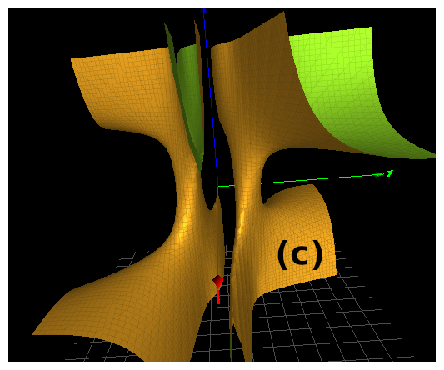}
\caption{The canonical component of character varieties for (a) the Whitehead link L5a1, (b) the Whitehead link sister L13n5885, (c) the Berg\'e link L6a2.}
\label{Canonical}
 \end{figure}

We refer to some torsion free subgroups of rank $1$ of Bianchi groups.
A Bianchi group $\Gamma_k=PSL(2,\mathcal{O}_k)<PSL(2,\mathbb{C})$ acts as a subset of orientation-preserving isometries of the $3$-dimensional hyperbolic space $\mathbb{H}_3$ with $\mathcal{O}_k$  the ring of integers of the imaginary quadratic field $\mathcal{I}=\mathbb{Q}(\sqrt{-k})$. A torsion-free subgroup $\Gamma_k(l)$ is the fundamental group $\pi_1$ of a $3$-manifold defined by a knot or a link such as the figure-of-eight knot [with $\Gamma_{-3}(12)$], the Whitehead link [with $\Gamma_{-1}(12)$] or the Borromean rings [with $\Gamma_{-1}(24)$]. See References \cite{Grunewald1993} and \cite{MPQGR2} for more cases. 

Looking at \cite[Table 1]{MPQGR2}, We see that both the  Whitehead link $L5a1$ and its sister $L13n5885$  (that have the same volume) are of the type $\Gamma_{-1}(12)$. They correspond to the $3$-manifolds ooct$01_{00001}$ and ooct$01_{00000}$, respectively. We are also interested with the links $L6a2=6_2^2$ (the Berg\'e link) and link $L6a1=6_3^2$ \cite{Harada2012} that are of the type $\Gamma_{-3}(24)$ and $\Gamma_{-7}(6)$, respectively. The latter link is related to $3$-manifold otet$04_{00001}$.

The four links have fundamental groups of rank $1$ as follows

\begin{eqnarray}
&\pi_1(S^3\setminus L5a1)=\left\langle a,b|ab^3a^2bAB^3A^2B\right\rangle, \nonumber \\
&\pi_1(S^3\setminus L13n5885)=\left\langle a,b|a^2bAb^2A^2BaB^2\right\rangle,\nonumber \\
&\pi_1(S^3\setminus L6a2)=\left\langle a,b|a^2bAb^2A^2BaB^2\right\rangle,\nonumber \\
&\pi_1(S^3\setminus L6a1)=\left\langle a,b|ab^3a^2b^2AB^3A^2B^2\right\rangle \nonumber. 
\end{eqnarray}

Remarkably, as for the Hopf link, we find that the cardinality structure of conjugacy classes of subgroups (card seq) of the fundamental group $\pi_1(S^3\setminus L)$ for the four links $L$ is invariant under the repetitive action of the Golden mean and the Silver mean substitution. This points out an unexpected relationship of rank $1$ Bianchi groups to aperiodicity.

The card seq of such $4$ groups is
\begin{eqnarray}
&\eta_d(\pi_1(S^3\setminus L5a1))=[1 ,3, 6, 17, 22, 79, 94, 412, 616, 1659,  2938, 10641,\cdots], \nonumber \\
&\eta_d(\pi_1(S^3\setminus L13n5885))=[1, 3, 5, 12, 19, 60, 44, 153, 221, 517,  632, 2223,\cdots],\nonumber \\
&\eta_d(\pi_1(S^3\setminus L6a2))=[1, 3, 4, 9, 24, 59, 71, 156, 262, 1208 \cdots],\nonumber \\
&\eta_d(\pi_1(S^3\setminus L6a1))=[1, 3, 7, 23, 28, 134, 184, 694, 1353, 3466\cdots].\nonumber 
\end{eqnarray}

Unlike the case of $\pi_1(L)$ for the Hopf link $L=$L2a1, with links of the Bianchi family, the Golden and Silver mean maps do not preserve the original group. Only the card seq of $\pi_1(L)$ is invariant for the above $4$ links.

We find that the three former links have a character variety with two components. The reducible component corresponds to the character variety of the Hopf link complement and, as described in the introduction, is associated to a degree $3$ Del Pezzo surface. The irreducible (or canonical component) is characterized below, see Table \ref{LinkCharVar} for a summary.

\small
\begin{table}[h]
\centering
\caption{ Character varieties of fundamental groups whose reducible representations are that of the Hopf link. Column 1 identifies the group as well as the corresponding link and $3$- or $4$-manifold. Column 2 is the name of the link or the relation it has to magic state quantum computing based on qutrits (QT) or two-qubits (2QB). Column 3 is  for the relation(s) of the two-generator fundamental groups.  When the link is not the Hopf link, column 4 is for the canonical component(s) of the representations and its (their) type as a surface in the $3$-dimensional projective space.} 
\label{LinkCharVar}
\begin{tabular}{c|c|c|c}
\hline
link $L$      & name      & rel(s) link group $\pi_1(L)$ & character variety $f(x,y,z)$ \\
\hline
L2a1 & Hopf & $[a,b]=abAB$ &   $f_H=xyz-x^2-y^2-z^2+4$\\
. &. & .&  deg $3$ Del Pezzo \\
$\Gamma_0(2)$, L7n1 & QT related & $[a,b^2]$ & $yf_H$ \\
$\Gamma_0(3)$, L6a3 & 2QB related & $[a,b^3]$ & $(y^2-1)f_H$ \\
\hline
&&&\\
$\Gamma_{-1}(12)$, L5a1 &\scriptsize{Whitehead} &$ab^3a^2bAB^3A^2B$ & $xy^2z-y^3-x^2y-xz+2y$ \\
$ooct01_{00001}$ &WL &. &conic bundle, $K_3$ type\\

\scriptsize{$\Gamma_{-1}(12)$, L13n5885} & \scriptsize{sister WL} & $a^2bAb^2A^2BaB^2$  &$x^2y^2-xyz-x^2+1$  \\
$ooct01_{00000}$  &. & . & deg $4$ Del Pezzo, $K_3$ type\\

$\Gamma_{-3}(24)$, L6a2 & Berg\'e & $ab^3a^2b^2AB^3A^2B^2$ &\scriptsize{$xy^3z-x^2y^2-y^4-xyz+3y^2-1$} \\
$otet04_{00001}$ &.&  . & conic bundle, general type \\

$\Gamma_{-7}(6)$, L6a1 &  &\scriptsize{$abABa^2BAb^3ABabA^2baB^3$} &undetermined  \\
\hline
&&&\\
$\tilde{E}_6$ & $IV^*$ & $a^3b^3,ab^2aBA^2B$ & $xy^3-y^2z-x^2-2xy+z+2,$\\
 .&. & . & $y^4-x^2z+xy-4y^2+z+2$\\
.&. & . & $K_3$ type \\
 \hline
\end{tabular}
\end{table}
\normalsize

Using Sage software \cite{CharVar} developed from Ref. \cite{Ashley2017}, the $SL_2(\mathbb{C})$ character variety for the $3$ links $L5a1$, $L13n5885$ and L6a2 factorizes as the product of two polynomials

$$f_H(x,y,z)f(x,y,z),$$

\noindent
where $f_H(x,y,z)$ is the character variety for the Hopf link complement  as obtained in Sec. \ref{HopfL}. The polynomial $f(x,y,z)$ consists of the irreducible $SL_2(\mathbb{C})$ representations of  $\pi_1(L)$. For the Whitehead link, one gets $f(x,y,z)=xy^2z-y^3-x^2y-xz+2y$. It is important to mention that the character variety for the group $\pi_1(L)$ depends on the selected Wirtinger representation. In \cite[Section 4.2]{Ashley2017}, the relation for the fundamental group of the Whitehead link complement is taken to be $abaB [A,B] ABAb [a,b]$ instead of the one obtained from SnapPy \cite{SnapPy} so that the canonical component of the character variety contains an extra third order term. 

Passing to the description of the surface $f(x,y,z)$ in the $3$-dimensional projective space as the homogeneous polynomial $\Sigma(x,y,z,t)=0$, the main algebraic properties of $\Sigma$ remain the same whatever the choice of the Wirtinger representation of $\pi_1(L)$. For the Whitehead link, we find that the surface $\Sigma$ is birationally equivalent to a conic bundle with a Kodaira dimension $0$. More precisely $\Sigma$ belongs to the family of $K_3$ surfaces.

Table \ref{Canonical} provides the canonical component of the character variety for the links L13n5885 and L6a2 whose algebraic description is a degree $4$ Del Pezzo surface of the $K_3$ family and a conic bundle of the general type, respectively. Unfortunately, we could not determine the character variety attached to the link L6a1.

\subsection{The $SL_2(\mathbb{C})$ character variety of singular fiber $IV^*=\tilde{E}_6$}

\begin{figure}[h]
\centering 
\vspace{-30pt}
\includegraphics[width=4cm]{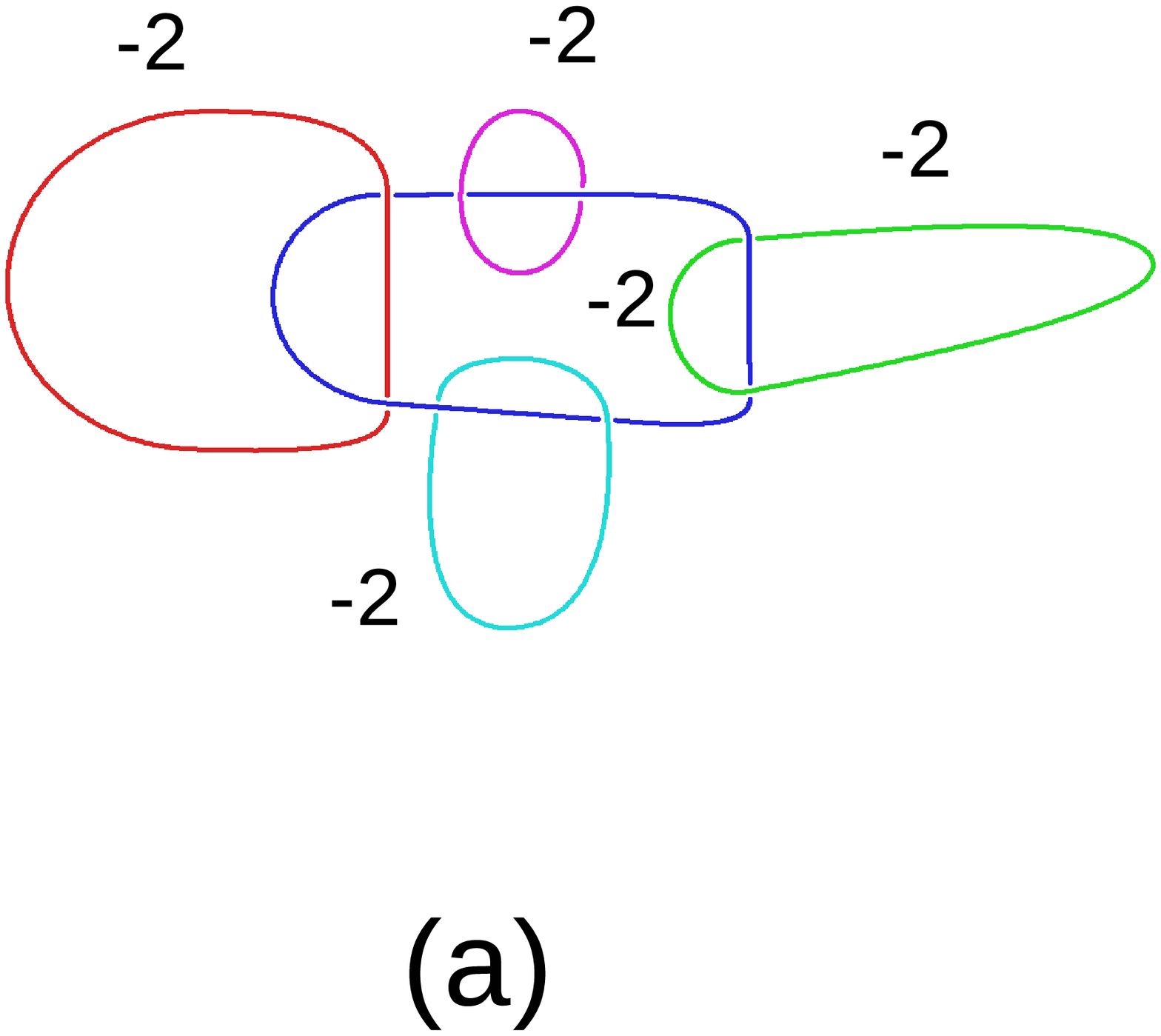}
\includegraphics[width=4cm]{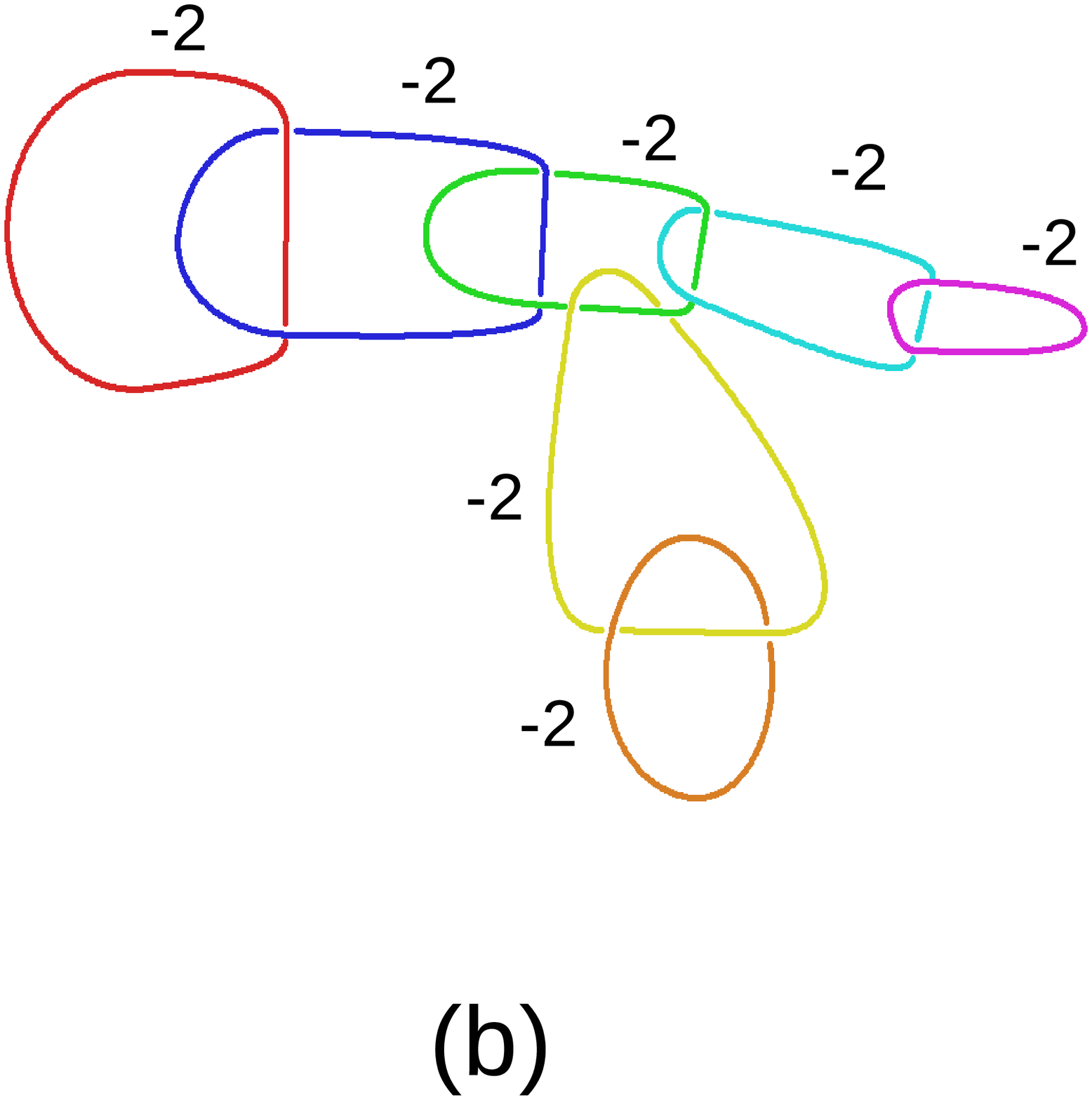}
\includegraphics[width=4cm]{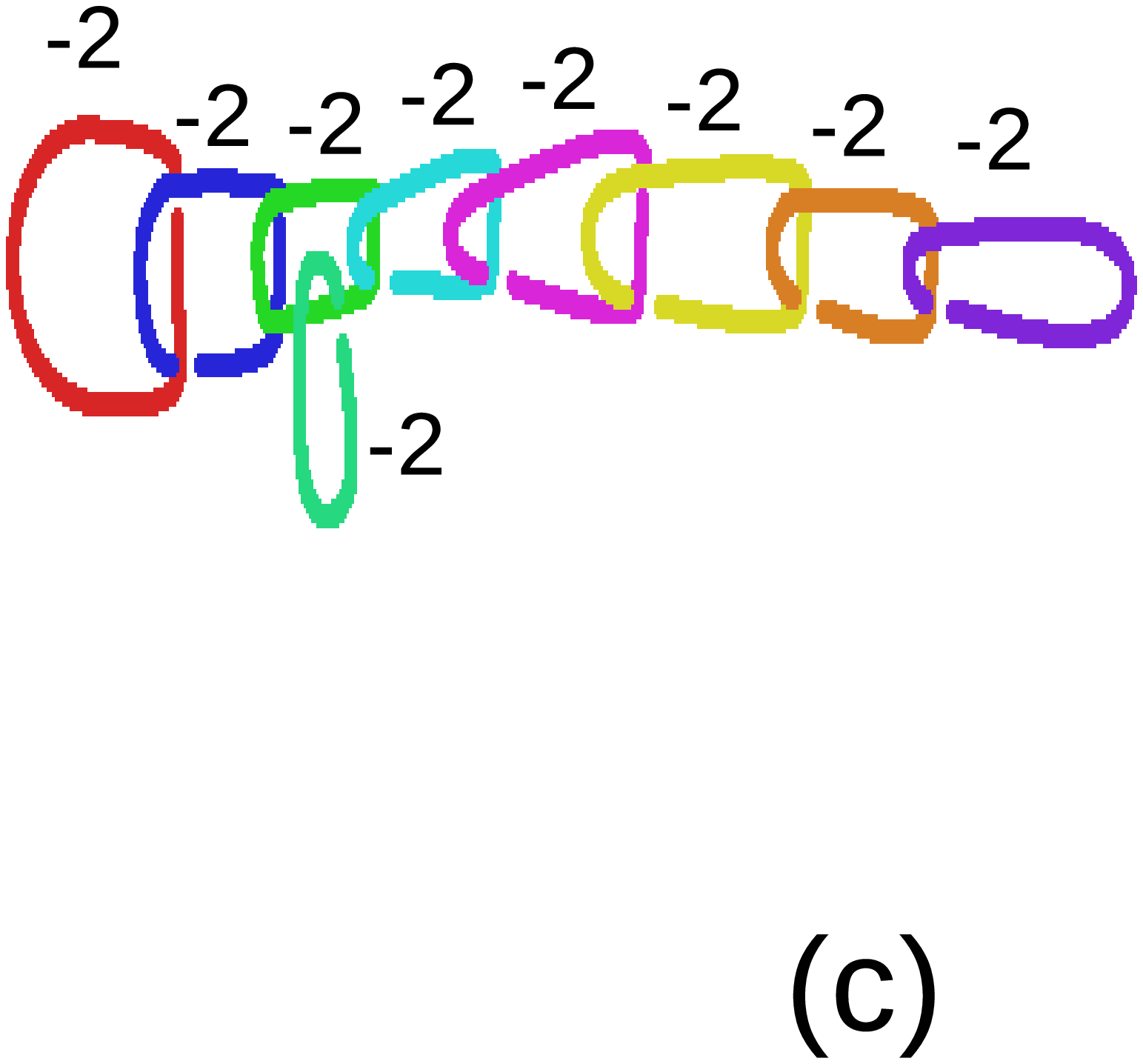}
\vspace{-18pt}
\caption{ A few singular fibers in Kodaira's classification of minimal elliptic surfaces. (\textbf{a}) Fiber $I_0^*$ (alias $\tilde{D}_4$), (\textbf{b}) fiber $IV^*$ (alias $\tilde{E}_6$), and (\textbf{c}) fiber $II^*$ (alias $\tilde{E}_8$). }
\label{E6E7E8tilde}
\end{figure}

In Reference \cite{PlanatQuantumReports}, we found connections between Kodaira singular fibers and magic state quantum computing.
The starting point of this viewpoint is the affine Coxeter-Dynkin diagram $\tilde{E}_8$ that corresponds to the fiber $II^*$ in Kodaira's classification of minimal elliptic surfaces \cite[p. 320]{Scorpian2005}, see Figure \ref{E6E7E8tilde}. Alternatively, one can see $\tilde{E}_8$ as the $0$-surgery on the trefoil knot $3_1$. The fundamental group of affine $E_8$ manifold has the card seq

\begin{equation}
\eta_d(\tilde{E}_8)=[1,1,{\bf 2},{\bf 2},1,~{\bf 5},{\bf 3},2,4,1,~1,12,{\bf 3},3,{\bf 4},~\ldots]
\end{equation}

\noindent
where the bold characters mean that at least one of the subgroups of the corresponding index leads to a MIC. The boundary of the manifold associated to $\tilde{E}_8$ is the Seifert fibered toroidal manifold~\cite{Wu2012}, denoted $\Sigma'$ in~\cite{MPQGR1} (Table 5).

For this sequence, the coverings are fundamental groups of \cite[p. 20]{PlanatQuantumReports}:

$$[\tilde{E}_8,\tilde{E}_6,\{\tilde{D}_4,\tilde{E}_8\},\{\tilde{E}_6,\tilde{E}_8\},\tilde{E}_8,
~\{BR_0,\tilde{D}_4,\tilde{E}_6\},\{\tilde{E}_8\},\{\tilde{E}_6\},\{\tilde{D}_4,\tilde{E}_8\},\tilde{E}_6,\cdots]$$

\noindent
The subgroups/coverings are fundamental groups for $\tilde{E}_8$ $\tilde{E}_6$, $\tilde{D}_4$, or $BR_0$, where $BR_0$ is the manifold obtained by zero-surgery on all circles of Borromean rings. 

One sees that the singular fiber $IV^*=\tilde{E}_6$ appears as the degree $2$ covering of $II^*=\tilde{E}_8$.  The fundamental group is 

\begin{equation}
\pi_1(S^4\setminus\tilde{E}_6)=\left\langle a,b|a^3b^3,ab^2aBA^2B\right\rangle,
\end{equation}

\noindent
where $S^4$ is the $4$-sphere.

We already found an invariance of the card seq of $\pi_1(L)$ under the Golden mean substitution (the Fibonacci map) or under the Silver mean substitution  
when $L$ is the Hopf link and when $\pi_1(L)$ is in the Bianchi family of $2$-generator groups. We now observe that this invariance is preserved when $L$ is the trefoil knot $3_1$, its surgery $\tilde{E}_8=3_1(0,1)$ and $\tilde{E}_6$. Aperiodicity is a feature of all the fundamental groups we encountered so far.


\begin{figure}[h]
\centering 
\vspace{30pt}
\includegraphics[width=6cm]{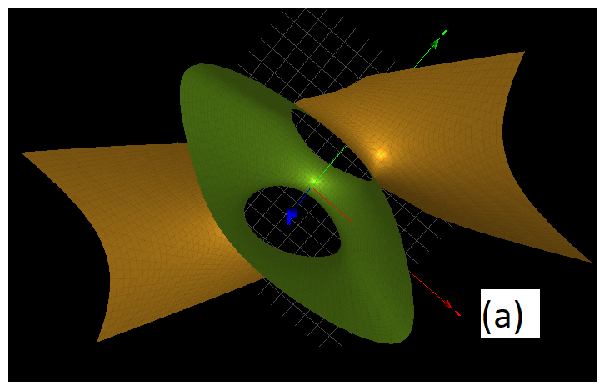}
\includegraphics[width=6cm]{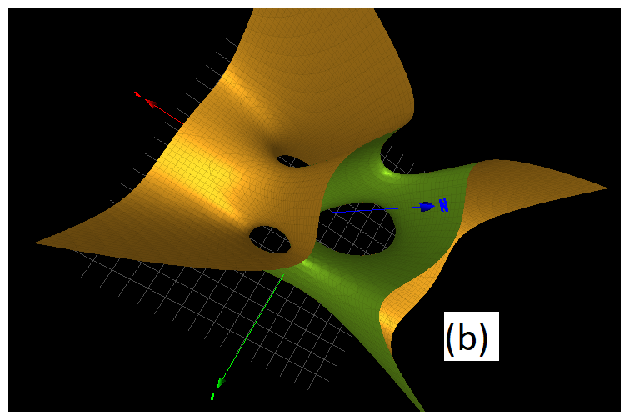}
\vspace{20pt}
\caption{The surfaces $f_1(x,y,z)$ and $f_2(x,y,z)$  in the character variety of singular fiber $IV^*=\tilde{E}_6$. Both surfaces are birationally equivalent to $K_3$ surfaces. }
\label{surfacesf1f2}
\end{figure}

Using Sage software \cite{CharVar}, the $SL_2(\mathbb{C})$ character variety of group $\pi_1(S^4\setminus\tilde{E}_6)$  factorizes as the polynomial product

$$f_H(x,y,z)(x-y)(xy-z+1)(x^2+xy+y^2-3)f_1(x,y,z)f_2(x,y,z),$$

\noindent
where $f_H(x,y,z)$ is the $SL_2(\mathbb{C})$ character variety for the fundamental group of Hopf link complement, $f_1(x,y,z)=xy^3-y^2z-x^2-2xy+z+2$ and $f_2(x,y,z)=y^4-x^2z+xy-4y^2+z+2$. A plot of the latter surfaces is in Figure \ref{surfacesf1f2}.
 
Passing to the description of the surfaces $f_1(x,y,z)$ and $f_2(x,y,z)$ in the $3$-dimensional projective space as $\Sigma_1(x,y,z,t)$ and $\Sigma_2(x,y,z,t)$ , one finds that $\Sigma_1$ is birationally equivalent to a conic bundle and $\Sigma_2$ to the projective plane $P^2$. Both surfaces shows a Kodaira dimension $0$ characteristic of $K_3$ surfaces.


The magic states from $\eta_d(\tilde{E}_8)$ at index $3$ and $4$ are $f_{QT}$ and $f_{2QB}$, as in Section \ref{magicQC} for the manifold L7n1 and L6a3, but their algebraic geometry is not that the Hopf link.
The associated $SL_2(\mathbb{C})$ character varieties are found to contain quadric surfaces $y-z^2+2$ (as for $\eta_d(\tilde{E}_8)$ itself) and $x^2+xy+y^2-3$, respectively.

Thus the existence of a magic state is not sufficient for the issue of topological quantum computing. The concept of $SL_2(\mathbb{C})$ character variety of the fundamental group has to be taken into account. In this respect, the affine $E_6$ manifold (the singular fiber $IV^*$) is a potential candidate.

\subsection{The $SL_2(\mathbb{C})$ character variety of singular fiber $I_0^*=\tilde{D}_4$}

Singular fibers occur inside a (minimal) elliptic fibration. Let us pass to the generic elliptic fiber $I_0$, a torus, and to the singular fiber $I_0^*=\tilde{D}_4$ shown in Figure \ref{E6E7E8tilde}a. A neighborhood of the singular fiber inside a $K_3$ surface leads to a plumbing diagram that is precisely $I_0^*$ \cite[Figure 3.15, p. 133]{Scorpian2005}.

As shown in the previous section, the link $\tilde{D}_4$ corresponds to the cyclic covering of degree $3$ of $\pi_1(\tilde{E}_8)$. The fundamental group is 

\begin{equation}
\pi_1(S^4\setminus\tilde{D}_4)=\left\langle a,b,c|a^2c^2,b^2c^2,aBCaBc\right\rangle.
\end{equation}

For this the $3$-generator group, the $SL_2(\mathbb{C})$ character variety is made of $7$ variable polynomials. Making use of the software available in \cite{CharVar}, it has the form

\begin{eqnarray}
&f(k,x,y,z,u,v,w)=(f_H(x,y,z)+wxk-2k^2) \nonumber \\
&(uk^2+vx-2u)(vk^2+ux-2v)(wk^2+xk-2W)(k^3+wx-2k) \nonumber \\
&(u^2-k^2)(uv-wk)(v^2-k^2)(uw-vk)(vw-uk)(w^2-k^2)(uy-2w)\nonumber \\
&(vy-2k)(wy-2u)(uz-2k)(vz-2w)(wz-2v)(yk-2v)(zk-2u),\nonumber 
\end{eqnarray}

\noindent
where $f_H(x,y,z)$ is the Hopf link polynomial of Section \ref{HopfL}.

Thus a section at constant $w$ and $k$ of the character variety for the link $\tilde{D}_4$ is simply a deformation of the character variety for the Hopf link, apart from trivial 
linear or quadratic polynomials.

As for the other links $L$ encountered so far, there exists an invariance of the card seq of $\pi_1(L)$ when $L=\tilde{D}_4$. Let us apply the map $a \rightarrow b$, $b \rightarrow abc$, $c \rightarrow a$ on the $3$ generators of $L$, the substitution map $T = \begin{pmatrix} 0 &1 &1\\ 1 &1 & 0 \\ 0 &1 & 0\end{pmatrix}$ is primitive since $T^3>>0$ and the Perron-Frobenius eigenvalue is the real root of the polynomial $\lambda^3-2 \lambda +1=0$, that is $\lambda_{PF} \sim 1.83928$, the Tribonacci constant \cite{Tribonacci}, see also \cite{BaakeBook} and \cite[Section 4]{CIMB2022} for mathematical details. This reveals the aperiodicity of the fundamental group.

\section{Conclusion}

We discovered connections between the $SL_2(\mathbb{C})$ character varieties for the fundamental groups of some links, the theory of algebraic surfaces and topological quantum computing. 
Our study was based on the Hopf link and links showing Hopf link character variety as a component. In particular, we were concerned with links in the Bianchi family (like the Whitehead link) 
and links for singular fibers in an elliptic fibration. The former define $3$-dimensional manifolds while the latter correspond to $4$-dimensional manifolds.

Our approach may connect to some theories of topological quantum field theory \cite{Kauffman} and quantum gravity \cite{Assanioussi}.

One starting point for future investigations may start from Reference \cite{Paluba2017} where several mathematical connections of character varieties to other branches of mathematics are proposed.
In particular, the Cayley's nodal cubic surface described as Equation (\ref{Fricke}) is in the family of smooth symmetric Fricke cubic surfaces \cite{Boalch2014}. The latter are isomorphic to a two parameter family of character varieties for the exceptional group $G_2(\mathbb{C})$. The group $G_2(\mathbb{C})$ arises as the group of automorphisms of the complex octonions, its (unique) semisimple conjugacy class is $6$-dimensional and relates to the $E_8$ root lattice thanks to the Fano plane representation of octonions. Three elements of $G_2(\mathbb{C})$, obtained from three lines passing through a single point in the Fano plane, generate a finite simple subgroup of $G_2(\mathbb{C})$ isomorphic to $G_2(2)'\cong U_3(3)$ of order $6048$.

Next, the group $U_3(3)$ stabilizes the split Cayley hexagon $GH(2,2)$ -- a $[63_3]$ configuration-- and its dual \cite[Table 8]{Planat2017}. The $63$ points of the hexagon may be encoded with three-qubit Pauli observables \cite{Levay2008}, the hexagon embeds $12096=2 \times 6048$ Mermin pentagrams (proofs of the Kochen-Specker theorem) which correspond to the number of automorphisms of $G_2(2)$ \cite{Planat2013} and finally the dual of the hexagon is obtained from the triple products of projectors defining the the Hoggar SIC-POVM (symmetric informationally complete-positive operator valued measure) \cite{Stacey2016}, \cite[Section 2.6 and Fig. 3]{PlanatGedik}.

\end{document}